\documentclass[a4paper,11pt]{article}
\pdfoutput=1 
 
\usepackage{jcappub} 
 
\usepackage[T1]{fontenc} 
 
\usepackage{color}
 
\title{Quantum contribution to luminosity of quasars}

\author[a]{J. E. Jacak,\note{Corresponding author.}}
 
\affiliation[a]{Department of Quantum Technologies, Wroc{\l}aw University of Science and Technology, Wyb. Wyspia\'nskiego 27, 50-370 Wroc{\l}aw, Poland}
 
\emailAdd{janusz.jacak@pwr.edu.pl}

\abstract{The accretion of galactic gas is regarded as the source of the giant luminosity of quasars. The gravitational energy converts itself into radiation close to the Schwarzschild horizon of the central supermassive black hole with efficiency of ca. 30 \% mass to radiation energy conversion rate. Particularities of such an extremely effective mechanism of mass to energy conversion are, however, still obscure. We propose to take into account quantum statistics properties of fermions, which could emit in close outer vicinity of the Schwarzschild zone a giant energy accumulated in the Fermi spheres of electrons and protons in degenerate quantum collective state created in this region by the gravitational compression of plasma. The release of photons is possible due to the local revoking of Pauli exclusion principle constraint induced by the rapid change of the homotopy of multiparticle trajectories beneath the innermost unstable circular orbit of the black hole, which causes the collapse of  Fermi spheres of electrons and protons.
}

\begin{document}
\maketitle
\flushbottom
 
\section{Introduction and motivation}

Quasars  are extremely luminous active galaxy nuclei powered by supermassive black holes with mass ranging from millions to tens of billions of Sun mass and consuming the gas from surrounding accretion disks. The effect is powerful, the radiant energy of quasars is giant and quasars have typically luminosities hundreds  of times greater than a galaxy such as the Milky Way. Quasar  surveys (over a million have been observed so far) have demonstrated that quasar activity was more common in the distant past -- the peak epoch was approximately 10 billion years ago. The closest known quasar is observed ca.  600 million light-years away from Earth. Record the farthest quasar is J0313-1806 observed in 2021
 with a 1.6 billion solar-mass black hole at z = 7.64, i.e., 670 million years after the Big Bang. Also in 2021, PSO J172.3556+18.7734 was detected as the most distant known radio-loud quasar discovered. A major concern is the vast amount of energy these objects would have to be radiating, if they were so distant. Commonly accepted explanation of such high power of quasars, that it is due to matter in an accretion disc falling into a supermassive black hole, which was  suggested in 1964 by Salpeter and Zeldovich \cite{salpet,zeld}. Dynamics of the accretion disc has been modelled by Shakura and Sunyaev in terms of classical fluid hydrodynamics and the release of  the gravitational field energy has been assessed due to by viscosity induced transport of angular momentum in opposite direction to accretion flow \cite{sunaev}. The model has been extended onto fully general gravitation approach by Novikov and Thorne \cite{novikov1} and was the subject of further intensive studies and more sophistical developments, e.g., towards rotating black holes in Kerr metric \cite{abramowiczg,reynolds}, however, still within classical hydrodynamic or magnetohydrodynamic approach.
 
 The energy transferred to the accretion disc matter on the cost of gravitational field of attracting central singularity and controlled by the angular momentum transport in opposite direction versus flow of matter in a flat disc is assumed to be eventually converted into electromagnetic radiation  by the thermal radiation including also the inverse Compton effect and bremsstrahlung. The realistic model of the active black hole luminosity has been formulated by Shapiro et al. \cite{shapiro}, who considered the accretion disc for microquasar Cignus X-1 powered by a 15 Sun mass black hole. By assumption the local temperature in the accretion disc ($10^9$ K for electrons and $10^{11}$ K for ions) the thermal radiation has been assessed as similar as observed, however, to elucidate the giant luminosity in X-ray range the Comptomization mechanism has been invoked, i.e., the increase of energy of soft thermal photons on the cost of energy of hot charge carriers in plasma. To assure the sufficient energy of these charge carriers the extremely high temperature in the accretion disc had to be assumed (just $10^{9-11}$ K). So hot plasma is assumed to occur in the small distance from the central black hole, at $r\sim 6 r_s$ (where $r_s$ is the event horizon radius of the black hole). Though the Shapiro model fits well to observations of microquasars, its generalization onto super-luminous giant quasars powered by $10^9$ Sun mass black holes or larger \cite{supereddington} is problematic, since the temperature $10^9 - 10^{11}$ K  of hot plasma in vast accretion disc of such large black holes  seems to be unrealistic. Moreover, the source for sufficient abundance  of soft photons in a small region where the Comptomization mechanism would be efficient is uncertain and  disputable.  Some numerical simulations done recently by Fragile et al. \cite{6p} allow to match to the observable luminosities of some not distant ($z<0.3$) active binary black hole objects \cite{farahzest}, but rather not to super-luminous remote quasars \cite{supereddington}. Higher radiation efficiency has been modelled within conventional classical magneto-hydrodynamic approach by inclusion of hypothetical giant magnetic component to accretion plasma in the case of a spinning black hole \cite{24p}, assuming, however, unrealistic extremal accretion mass rate to gain sufficiently large luminosity.
 
Models of the accretion disc  utilize the energy conservation condition and the transport of angular momentum in the flow of matter in gravitational potential of a black hole within a conventional hydrodynamic turbulence picture with some phenomenological assumptions regarding viscosity, inner disc edge and channels of energy dissipation imposed \cite{sunaev,merloni,novikov1,abramowiczg,shapiro}. In particular, the conventional models of the accretion assume the termination of the disc radiation at the inner disc edge located not closer to the singularity than at the innermost stable circular orbit (at $3r_s$, though the active region is not closer than $6r_s$ \cite{shapiro}), neglecting the nearest neighbourhood of the event horizon.  Such classical hydrodynamic or magnetohydrodynamic  models  are not applicable in closer vicinity of the event horizon \cite{ksiazka}. At extreme matter influx rates as in super-luminous quasars, these models have problems with fitting the observations and additional channels of gravitational energy to radiation transfer are still being sought.
 To explain the enormously intensive hard radiation of super-luminous quasars (despite its softening by gravitation relativity effect due to Doppler boosting and gravitational redshift), some other mechanism
 besides thermal radiation or other conventional mechanisms (Comptomization and bremsstrahlung) seems to be contributing.
 
 In the present paper we propose such a new channel of gravitational energy to radiation transfer close to the event horizon of the central black hole, i.e., in the region neglected in classical hydrodynamic-type previous models. We will demonstrate that the quantum in nature process undergoing on the rim of black hole photon sphere (at distance $1.5r_s$ from the gravitational singularity) can substantially supplement the overall luminosity of quasars and in the case of an uppermost rate of mass consumption by the black hole, it can help to match the observable   radiation energy/mass conversion rate on the level of 30 \% for super-luminous active gravitation centres.
Such high radiation energy/mass conversion rate is actually needed to explain the typical luminosity of giant quasars being of order of $10^{40}$ W and powered by ca. billion Sun mass black holes consuming of order of $10$ Sun mass per year (i.e., $0.1$  Earth mass per second), in extreme case of $1000$ Sun mass per year ($10$ Earth mass per second). Such rates of mass consumption are inferred from the comparison of 
central black holes in quasars, which  vary between $ 10^5 - 10^9$ of solar masses, as  have been measured using a reverberation mapping, and cannot increase during the quasar activity period beyond an average  mass of closer non-active super-massive black holes. Several dozen nearby large galaxies, including our own Milky Way galaxy, that do not have an active centre and do not show any activity similar to a quasar, are confirmed to contain a similar supermassive black hole in their  centres. Thus it is now thought that all large galaxies have a giant black hole of this kind, but only a small fraction have sufficient matter in the right kind of orbit at their centre to become active and power radiation in such a way as to be seen as quasars. However, if quasars are active for a long time and their mass cannot increase too much, the efficiency of converting matter into energy must be high to explain the gigantic luminosity of  quasars.

An argument supporting a model of quasar luminosity from the accretion disc close to the Schwarzschild horizon of a central black hole  is the observed variation in time of quasar activity.
Quasars' luminosities vary with time scale that ranges from days to hours. This means that quasars generate and emit their energy from a very small region, since each part of the quasar would have to be in contact with other parts on such a time scale as to allow the coordination of the luminosity variations. This would mean that a quasar cannot be larger than a few light hours or days across -- just as the size of the Schwarzschild horizon radius of the central supermassive black holes, $r_s=\frac{2GM}{c^2}$, where $M$ is the mass of the black hole, $G$ is the gravitational constant, $c$ is light velocity in the vacuum. For exemplary supermassive black holes such radius is 0.08 au for Sigittarius A* in Milky Way with mass $M=8.2 \times 10^{36}$ kg (4.15 million $M_\odot$), or 1300 au for largest known black hole TON 618 with mass $M=1.3 \times 10^{41}$ kg (66 billion of $M_\odot$).  The emission of a large amount of power from a small region requires a power source far more efficient than the nuclear fusion that powers stars. The conversion rate of the gravitational potential energy to the radiation energy at the mass influx to a black hole reaches ca. 30 \%   in quasars, compared to 0.7 \% for the conversion of mass to energy in a star like our Sun. It is the only known process that can produce such high power over a very long term. Stellar explosions such as supernovas and gamma-ray bursts, and direct matter–antimatter annihilation, can also produce very high power output, but supernovas only last for days, and the universe does not appear to have had large amounts of antimatter at the relevant times. A physical mechanism of so effective mass-energy conversion in quasars is still obscure to some extent, as the thermal radiation of hot ionized gas, bremsstrahlung, synchrotron radiation and the inverse Compton effect  cannot be so efficient at case of super-luminous quasars \cite{supereddington} with vast accretion discs probably not so hot as in the case of more concentrated and smaller centres (like Cignus X-1 \cite{shapiro}).
 Probably some additional mechanism of energy conversion  contributes in the vicinity of the Schwarzschild  horizon to produce an additional amount of radiation in  quasars.  We propose such a mechanism taking into account collective quantum effects in dense fermion plasma associated with a phase transition in multiparticle systems passing the rim of the photon sphere of a black hole.
 
The paper is organized as follows. In the next paragraph we propose the new mechanism of gravitational energy conversion into e-m radiation, which has a quantum character and is effective at passing the photon sphere rim of the black hole (at the distance $1.5 r_s$ from the origin with gravitational singularity). The physical details of the proposed mechanism are next explained. The comparison with observations are placed in the next paragraph where the new model is suggested also to supplement current concept of high energy radiation at collapse of neutron star  merger exceeding the Tolman-Oppenheimer-Volkoff limit and, on the other hand, some short time-lasting luminosity changes of active galactic nuclei (as recently observed for AGN 1ES 1927+654 \cite{ga1flare}). Some numerical estimations and additional comments and mathematical proofs are shifted to Appendices.
 
\section{Proposition of quantum mechanism driving quasar activity below the innermost unstable circular orbit of a black hole}
 
According to the conventional models of the accretion disc (\cite{sunaev,novikov1} and developments e.g.,  \cite{abramowiczg,reynolds,merloni,ksiazka}), the radiation of such disc terminates at the inner edge of the disc not closer to the event horizon than ca. $6r_s$ \cite{shapiro}. Such a common theoretical assumption displays the fact that the classical hydrodynamic approach on which  the theory of the accretion disc is based on \cite{sunaev,ksiazka}, does not apply to the region closest to Schwarzschild horizon where the conventional scattering regime of particles in plasma is substituted by another regime  resembling some-kind of laminarization of matter movement induced by the giant gravity, prevailing and marginalizing  local interparticle effects the closer to the horizon.
 
The timescale of more or less regular changes in the brightness
of  quasars suggest that the responsible radiation  source  must be concentrated close to
the event horizon rather than located in remote regions of the
accretion disk (or in distant jets). 
This hypothetical source of some part of quasar radiation located close to the event horizon must be governed by completely different mechanism than that already  described within classical  hydrodynamic accretion  models applicable in more remote regions \cite{sunaev,novikov1,shapiro,ksiazka}.
 
We propose some supplementation to a model of quasar luminosity source including the region just below the innermost unstable circular orbit (the rim of the photon sphere in Shwarzschild metric), taking advantage of the specific change of homotopy of trajectories of particles which if they enter the photon sphere, they irrevocably spiral onto the Schwarzschild horizon in the laminarized manner governed by the central gravitational singularity no matter how large the initial energy and angular momentum of  particles are  and no matter how strongly they mutually interact.   This is the result of spacetime curvature induced by the gravitational singularity at small distances, what can be  illustrated in standard  Schwarzschild metric. Sharp changing of the homotopy of classical admissible trajectories for particles passing the innermost unstable circular orbit (which coincides with the photon sphere rim for non-rotating and uncharged black hole at $r=1.5 r_s$ distance from the central singularity) causes the quantum phase transition consisting in the local repeal of the Pauli exclusion principle, which results in the rapid collapse of Fermi spheres of collective systems of electrons and protons in infalling plasma in the accretion disk. The local density of fermions in close vicinity of the event horizon attains large values due to the gravitational compression. Fermi liquids of electrons and protons are quantumly degenerated here and accumulate great energy in their Fermi spheres on the cost of the gravitational energy.
The conversion of the gravitational field energy into electromagnetic radiation at rapid collapse of Fermi spheres of fermions  is  very effective for sufficiently high scale of transport of the mass and its local compression close to the event horizon, and can help to explain the observable luminosity of quasars. The consumption of the mass by the black hole cannot exceed in this model the limit imposed by the uppermost density of matter at the photon sphere rim (of order of the density of hadrons in atom nucleus or  the density of neutrons in neutron star at  Tolman-Oppenheimer-Volkoff limit). This constraint regulates the matter consumption rate in dependence of the black hole mass in the case of not limited abundance of matter supply to the accretion disc from surroundings.  The phenomenon of suggested Fermi sphere decay has a topological character sourced in the homotopy of Schwarzschild geodesics in curved spacetime  and displays a unique situation when the energy of fermions accumulated  in Fermi spheres of electrons and protons  in dense highly compressed plasma can be released due to quantum phase transition related to the local decay of quantum statistics triggered  by the  change of the homotopy of trajectories close to central singularity of  the black hole.
At sufficiently high number of particles compressed to a small volume by the  giant gravitation close to the event horizon, just at the innermost unstable circular orbit at $r=1.5r_s$ (coinciding with the photon sphere rim),  the mass to energy  conversion rate at the collapse of the Fermi spheres of electrons and protons reaches the value of 30 \% provided that the density of matter at the rim of the photon sphere attains its uppermost limit. Such a mechanism seems to be responsible for a significant contribution to the luminosity of giant super-luminous quasars for which the matter supply from their surroundings is limited only by the uppermost density limit at the photon sphere rim. The detailed proof of the proposed model
is demonstrated in the following paragraphs.
 
\subsection{Change of trajectory homotopy below the innermost unstable circular orbit and the collapse of the Fermi sphere as the radiation source} 
 
If identical indistinguishable particles distributed in some  space (mathematically defined as a manifold $M$) can exchange their positions in $M$ (i.e., if the classical trajectories for such exchanges are admissible in the manifold $M$), the quantum statistics of these particles is assigned by a scalar unitary representation of the braid group \cite{mermin1979aaa,lwitt} (as shortly presented in  Appendix \ref{a}). The braid group is defined for $N$ identical indistinguishable particles on the manifold $M$ as the first homotopy group of the multiparticle coordination space of this system, i.e., the braid group is equal to $\pi_1(F_N)$, the first homotopy group of the space $F_N$, where $F_N=(M^N-\Delta)/S_N$ is the classical coordination space of $N$ identical indistinguishable particles in $M$ ($\Delta$ is the diagonal subset of the $N$-fold product of the manifold $M^N=M\times\dots \times M$ and subtracted from $M^N$ to assure the particle number conservation, whereas the division by the permutation group $S_N$ introduces particle indistinguishability -- points in $F_N$ which differ by only numbering of particles are unified). Various scalar unitary representations of $\pi_1(F_N)$ assign different quantum statistics of the same classical particles (more specific explanations are shifted to
 Appendix  \ref{a}).
 
In 4D spacetime only two scalar unitary representations of the braid group are possible, corresponding to bosons and fermions \cite{mermin1979aaa}. The most prominent consequence of quantum statistics is the Pauli exclusion principle, which   asserts that quantum particles of fermionic type  cannot share any common single-particle quantum state. In particular, fermions cannot approach a spatial region already occupied by another fermion, and thus they mutually repulse themselves. This is called the quantum degeneracy repulsion and the related  pressure is the origin of stopping the collapse of white dwarfs or neutron stars. In the former case the degeneracy pressure of electrons plays the role \cite{chandrasekhar}, whereas in the latter case of neutrons \cite{tolman,volkoff,olandau}.
 
Pauli exclusion principle leads to the formation of the so-called Fermi sphere in the case of large number of identical fermions in some volume, when the chemical potential $\mu$ (the increase of total system energy in the result of the addition of a single fermion to a multiparticle system) is much greater than the temperature in the system  in energy scale, $k_BT$, where $k_B$ is the Boltzmann constant. In such a case, referred to as quantum degenerated Fermi system,  particles are forced to occupy  consecutive in energy single-particle states one by one (these states can be numbered by the momentum in macroscopic systems, because the momentum  is a good quantum number at translational symmetry, and also for macroscopic finite but large systems with local translational symmetry and conventional periodic Born-Karman boundary conditions imposed \cite{landau1972}). The  occupation one-by-one of ordered in energy single-particle states results in a great accumulation of energy in the  Fermi sphere, and the accumulated energy can achieve a giant value in dense systems. In finite size macroscopic systems with spatial volume $V$, the momentum $\mathbf{p}$ of single-particle states is discrete, and according to periodic Born-Karman boundary conditions the density of quantum states in the phase space is  $\frac{dVd^3\mathbf{p}}{(2\pi\hbar)^3}$ (in agreement with the Bohr-Sommerfeld quasiclassical quantization rule, on the other hand \cite{landau1972}), $\hbar=1.05 \times 10^{-34}$ Js is the reduced Planck constant. Thus, $N$ fermionic particles  fill the Fermi sphere in momentum space up to some radius  called as the Fermi momentum separating discrete filled states inside  the sphere from outside  empty ones (at ground state of the total system, i.e., at $T=0$ K and provided that the single-particle energy spectrum is isotropic in momentum space, as e.g., for free  particles either relativistic with energy $\varepsilon(\mathbf{p})=\sqrt{p^2c^2+m^2c^4}-mc^2$ or classical ones with energy $\varepsilon(\mathbf{p})=\frac{p^2}{2m}$). This picture is maintained even in nonzero temperature, provided that the chemical potential of fermions $\mu\gg k_BT$ (where the chemical potential $\mu$ is the energy increase when a single particle is added to the system). This condition is easy to be satisfied in sufficiently dense systems, for example in astrophysical problems considered in this paper the Fermi sphere is almost the same as at $T=0$ K at temperatures even of order of $10^9$ K (the proof is placed in paragraph \ref{efficiency}). 
 
As an elementary illustrative example let us consider  free electrons in a normal metal with the typical concentration of order of $10^{23}$ (of order of Avogadro number) per cm$^3$. These electrons constitute the large Fermi sphere with Fermi radius in momentum space $p_F \simeq 1.5 \times 10^{-24} $ kg m/s and with accumulated large total energy $\sim 3\times10^{10}$ J/m$^3$. This energy cannot be released because all fermions are blocked in their single-particle stationary states by Pauli exclusion principle, i.e., all lower single-particle states in the Fermi sphere are occupied and thus there is no room for fermions to jump  from the higher energy states to  lower ones. In this example of normal metal the energy of upper electrons in the Fermi sphere, $\frac{p_F^2}{2m}$ ($m=9.1 \times 10^{-31}$ kg is the rest mass of electron) at $T=0$ K reaches $90000$ K  (in thermal scale, i.e., in units for which the Boltzmann constant $k_B =1$). This Fermi energy is the chemical potential at $T=0$ K and it weakly changes with the temperature up to  melting temperature of a metal. So hot electrons on the Fermi surface even at $T=0$ K do not melt the metal by themselves, as they cannot give back their energy blocked in the Fermi sphere by the Pauli exclusion principle.
 
The Fermi momentum in the degenerate quantum liquid of arbitrary fermions depends solely  on particle concentration,
\begin{equation}
	\label{fm}
	p_F=\hbar(3\pi^2 \rho)^{1/3},
\end{equation}
where $\hbar$ is the reduced Planck constant and $\rho=\frac{n}{V}$ is the concentration of fermions, i.e., $n$ is the number of particles in the spatial volume $V$.
The Fermi momentum is independent of interaction of fermions according to Luttinger theorem \cite{luttinger}.   This follows from the fact that the phase space volume  $V \frac{4}{3} \pi p_F^3$ with the position space volume $V$ and the volume of momentum-space  of spherical shape $\frac{4}{3}\pi p_F^3$, corresponds to $  n=2(V\frac{4}{3}\pi p_F^3)/ h^3$ quantum states according to the  Bohr-Sommerfeld rule \cite{landau1972} ($h=2\pi\hbar$ is the Planck constant and factor  $2$ accounts here for the additional  spin degree of freedom  of fermions). The formula for Fermi momentum, as quasiclassically derived, is independent of interaction of fermions in Fermi liquid even with arbitrary strong particle interaction. With Fermi momentum the Fermi energy is linked -- the single-particle energy of a particle with Fermi momentum (equal to the chemical potential of fermions in multiparticle system at $T=0$ K). Thus for electrons in a metal with  $p_F\simeq 10^{-24}$ kg m/s the Fermi energy (in gas approximation) $\varepsilon_F=\frac{p_F^2}{2m} \simeq 1.2 \times 10^{-18} $ J (or $\sim 90000$ K in thermal units).
 
The whole Fermi sphere collects the energy of all fermions from the bottom of this sphere to its surface and this energy  per spatial  volume $V$ (neglecting interaction of fermions) equals to
\begin{equation}
	\label{en}
	\begin{array}{l}
		E=\sum_{\mathbf{p}} \varepsilon(\mathbf{p}) f(\varepsilon(\mathbf{p}))\\
		= \frac{V}{(2\pi \hbar)^3}\int d^3\mathbf{p}\varepsilon(\mathbf{p})f(\varepsilon(\mathbf{p}))\\
		=\int_0^{p_F}dp \int_0^{\pi} d \theta\int_0^{2\pi} d\phi  p^2 sin\theta \varepsilon(\mathbf{p}) \frac{V}{(2\pi \hbar)^3}\\
		=\frac{V}{2 \pi^2 \hbar^3}\int_0^{p_F} dp p^2 \varepsilon(p),\\
	\end{array}    	
\end{equation}
where the sum runs  over occupied states only, what is guaranteed by Fermi-Dirac distribution function $f(\varepsilon(\mathbf{p}))=\frac{1}{e^{(\varepsilon(\mathbf{p})-\mu)/k_BT}+1}\rightarrow_{T\rightarrow 0}1-\Theta(\varepsilon(\mathbf{p})-\varepsilon_F)$ (here $\Theta(x)$ is the Heaviside step function and $\varepsilon_F=\mu(T=0)$),  $p,\theta, \phi$ are spherical variables in momentum space and $\varepsilon(\mathbf{p})$ is the kinetical energy of a fermion equal to $\frac{p^2}{2m}$ (in nonrealtivistic case), $\sqrt{p^2c^2+m^2c^4}-mc^2$ (in relativistic case) or $cp$ (in ultrarelativistic case). The factor $\frac{V}{(2\pi \hbar)^3}$ is the density of quantum states, i.e., the number of single-particle quantum states in the element of the phase space $Vd^3\mathbf{p}$ -- this is  the density of states $\frac{dVd^3\mathbf{p}}{(2\pi\hbar)^3}$ integrated over the whole spatial volume $V$ of the system, because the energy is independent of spatial position.
We thus see that in the Fermi sphere the energy can be stored up to a giant value (the larger the higher Fermi momentum is; in normal metals of order of $10^{10}$ J/m$^3$).
This giant energy cannot be released as electrons cannot lower their energy, because all preceding states with lower energy are already occupied.
 
The energy reservoir in the Fermi sphere can be even greater than in metals because it depends on density of matter via the Fermi momentum given by Eq. (\ref{fm}) and the concentration of fermions can be larger than that of free collective  electrons in metals. In another example -- a neutron star with the density of order of $10^{18}$ kg/m$^3$ (i.e., of the order of two Sun mass compressed to the compact neutron star with radius of ca. $10$ km), the concentration of neutrons is of order of $10^{45}$ 1/m$^3$. For such a concentration of neutrons  the neutron Fermi sphere energy in the whole star attains the range of $10^{47}$ J, just as the energy of frequently observed cosmic short giant gamma-ray bursts (assuming the isotropy of their sources) -- for some more particularities of this example cf. Table \ref{tab-kw1}.
 
\begin{table*}
	\caption{Fermi momentum $p_F$ (acc. to Eq. (\ref{fm})),  Fermi energy $\varepsilon_F=\varepsilon(p_F)$ (for relativistic case of kinetic energy $\varepsilon(\mathbf{p})=\sqrt{c^2p^2+m_n^2c^4}-m_nc^2$, $m_n=1.675\times 10^{-27}$ kg -- the mass of neutron) and total energy of the Fermi sphere $E$ (acc. to Eq. (\ref{en})) released at the decay of the statistics for the example of density $\xi$ and radius $r$ of a neutron star  ($n$ total number of neutrons in the star). The collapse of Fermi sphere in a neutron star caused by the gravitation induced specific homotopy of trajectories near the Schwarzschild radius, which precludes particle exchanges and locally revokes the Pauli exclusion principle, is a good candidate for isotropic source of short giant gamma-ray burst with observable energy $\sim 10^{47}$ J.}             
	\label{tab-kw1}  	
	\centering          	
	\begin{tabular}{c c c c c c}    	
		\hline\hline 	
		$\xi$ [kg/m$^3$]  & $r$ [km] & $n$& $p_F$ [kg m/s]& $\varepsilon_F$ [GeV] & $E$ [J] \\	
		\hline        	
		$5\times10^{18}$& $10$ & $1.13\times 10^{58}$&$4.52 \times 10^{-19}$ &$0.32$ & $1.84 \times 10^{47}$ \\
			\hline        	
		$2 \times 10^{18}$&10&$4.7 \times 10^{57}$&$3.37 \times 10^{-19}$&$0.2$&$4.8 \times 10^{46} $\\
		\hline
		$1.0 \times 10^{19}$&$8$&$1.08 \times 10^{58}$&$5.57\times 10^{-19}$&$0.58$&$3 \times 10^{47}$\\
		\hline
		$2.5 \times 10^{18}$&$8$&$2.97 \times 10^{57}$&$3.62 \times 10^{-19}$&$0.24$&$3.5 \times 10^{46}$\\
		\hline        	
		
	\end{tabular}
\end{table*}
 
The similar energy estimation can be  performed  for any quantumly degenerated fermion system, i.e., when the chemical potential -- the energy increase caused by the addition of a single particle to multiparticle  system -- is much greater than the actual thermal energy in the system, $k_B T$, $k_B=1.38 \times 10^{-23}$ J/K is the Boltzmann constant, $T$ is the absolute temperature in the system (in normal  metal this chemical potential  is of order of 10 eV, or $10^5$ K [in units with $k_B=1$] and thus $\mu$ is practically equal to the Fermi energy $\varepsilon_F$, the upper electron energy at $T=0$ K, i.e.,  $\mu\simeq \varepsilon_F\gg k_BT$ even at melting temperature.
 
If one considers an accretion disc of a quasar, then the local density of electrons and protons (assuming accretion of neutral hydrogen) grows with falling of the matter towards the Schwarzschild horizon. The originally diluted neutral gas (let's say of a hydrogen cloud)  ionizes itself due to friction in the accretion disc and eventually becomes a degenerate two-component Fermi liquid of electrons and protons  despite the high temperature. Both these Fermi liquids  attain an ultra-high concentration in an increasingly flattened and compressed region in vicinity of the event horizon. For two component plasma,  both Fermi spheres  of electrons and protons contribute to the energy storage. At the same concentration of electrons and protons (due to the neutrality condition of plasma in the disc) electrons accumulate larger kinetic energy than protons because of lighter mass (the relativistic kinetic energy of the electron and  proton is $\varepsilon(\mathbf{p})=\sqrt{c^2p^2 +m^2c^4}-mc^2$ with $m=m_e=9.1 \times 10^{-31} $ kg and $m=m_p=1.67 \times 10^{-27}$ kg, respectively).
 
The energy accumulated in the Fermi spheres of electrons and protons can be released in the form of the electromagnetic radiation if the Pauli exclusion principle is locally waived  in close vicinity of the event horizon, due to local decay of quantum statics (as proved in Appendix C). Charged particles couple to an electromagnetic field and the collapse of their Fermi spheres is accompanied with the emission of photons in agreement with Fermi golden rule for quantum transitions between initial states of particles in the Fermi sphere and their ground state. 
 
\subsection{Energy efficiency of the collapse of Fermi spheres of electrons and protons in the accretion disc near the  event horizon of a quasar}
 
For concreteness of the estimation let us assume that the central black hole in quasar consumes 5.6 $M_{\odot}$ per year, i.e., ca.  $0.06$ Earth mass per second. Let us assume the stable uniform in time process of matter accretion. The transport of matter across the disk is steady,  thus we can perform calculation e.g., per a single second. Using Eqs (\ref{fm}) and (\ref{en}) one can assess the energy stored in the Fermi spheres for electrons and protons, if all the electrons and protons from the gas mass equalled to $0.06 $ Earth mass, are compressed to the spatial volume $V$  per second (this volume depends on the distance from the event horizon). The local Fermi momentum
\begin{equation}
	\label{fmm}
	p_F(r)=\hbar(3\pi^2 \rho(r))^{1/3}=\hbar\left(3\pi^2 \frac{n}{V(r)}\right)^{1/3},
\end{equation}
where $r$ is the distance from the centre. $p_F(r)$  is constant in time and grows across the disk with increasing local concentration $\rho(r)=\frac{dn}{dV}=\frac{n}{V(r)}$, the same for electrons and protons. The latter equality holds for steady accretion and $n$ is the total number of electrons (or protons) per second, compressed in total to the volume $V(r)$ at the distance $r$ from the origin with central gravitational singularity. This means that  portions $dn$ of electrons and protons in infinitely small consecutive periods $dt$ incoming in radial direction towards the central singularity compressed to $dV(r)$ at radius $r$ add up in due of a single second time period to the total constant flow of mass (in the example, of $0.06 \times M_Z$ kg/s, the Earth mass $M_Z=5.97 \times 10^{24}$ kg) and as the whole are compressed locally at $r$ to $V(r)$. The locally accumulated energy in the Fermi spheres of electrons and protons  grows with lowering $r$ due to the increase of the compression caused by the gravitational field. This energy is proportional to $V(r)$ and, moreover, depends on $V(r)$ via the local  Fermi momentum  (\ref{fmm}) and in accordance  with  Eq. (\ref{en})) can be expressed as,
\begin{equation}
	\label{enn}
	\begin{array}{l}
		E(r)=E_e(r)+E_p(r),\\
		E_e(r)=\frac{V(r)}{2 \pi^2 \hbar^3}\int_o^{p_F(r)}dp p^2 \left( \sqrt{p^2 c^2 +m_e^2 c^4}-m_e c^2\right),\\
		E_p(r)=\frac{V(r)}{2 \pi^2 \hbar^3}\int_o^{p_F(r)}dp p^2 \left(\sqrt{p^2 c^2 +m_p^2 c^4}-m_p c^2\right),\\
	\end{array}
\end{equation}
where the energy $E_{e(p)}$ refers to electrons (protons).

At  the critical radius $r^*$ close to Schwarzschild zone (we argue that $r^*=1.5r_s$ as detailed in Appendix \ref{b})  the decay of  quantum statistics takes place due to  the topology reason (cf. Appendices \ref{a} and \ref{b}) and both Fermi spheres of electrons and protons collapse. The amount of energy given by Eq. (\ref{enn}) per one second can be thus released in the vicinity of the Schwarzschild horizon (at the rim of the photon sphere). Those energies continuously released add up per single second to ca. $10^{40}$ J and can contribute in large part to the observed luminosity of quasar (of order of just $10^{40}$ W). This process  undergoes by portions $dn$ of particle flow incoming to $r^*$ region in infinite small time periods $dt$, adding up   to  $0.06 \times  M_Z$ kg per second in total. The value  of the  released energy depends on local  Fermi momentum and  attains  $10^{40}$ J at sufficiently high level of compression, i.e., at sufficiently small $V(r^*)$ determined from the self-consistent system of Eqs (\ref{fmm}) and (\ref{enn}) if one assumes $E(r^*)=10^{40}$ J. Obtained in this way $V(r^*)$ allows for the assessment of local matter density -- it occurs of order of the uppermost limit for matter density as in neutron star at Tolman-Oppenheimer-Volkoff limit (i.e., of order of density of atomic nuclei), which evidences self-consistency of the model.

To the initial mass of a gas (assuming to be composed of hydrogen $H$)
contribute mostly protons (ca. 2000 times more massive than electrons), thus the total number of electrons, the same as the number of protons falling onto the considered black hole, equals to, $n\simeq 0.06 M_Z /m_p\simeq 2.14 \times 10^{50}$ per second. Simultaneously solving Eqs (\ref{fmm}) and (\ref{enn}), assuming $n=2.14 \times  10^{50}$  in volume $V(r^*)$ and released energy $E(r^*)=10^{40}$ J,  we find the volume of plasma compression $V(r^*)=0.5 \times 10^5$ m$^3$ and electron or proton Fermi sphere radius $p_F(r^*)=5.4 \times 10^{-19}$ kg m/s. Electrons and protons (their amount per second) are compressed to the same volume $V(r^*)$ (due to the neutrality of plasma), hence,  their  concentration at $r^*$, $\rho(r^*)=4.3 \times 10^{45}$ 1/m$^3$. The mass density at $r^*$ (including mass equivalent to the energy stored up in Fermi spheres of electrons and protons) is thus  $\xi(r^*)=\frac{0.06 M_Z}{V(r^*)}+\frac{E(r^*)}{c^2 V(r^*)}\simeq 9 \times 10^{18}$ kg/m$^3$ (similar to mass density in neutron stars). The  released energy of $E(r^*)=10^{40}$ J is equivalent to 30\% of the falling mass of $0.06$ Earth mass (per second). It means that the compressed plasma with degenerate Fermi liquid of electrons (and also of protons) is at $r=r^*$ by 30\% more massive than initial remote diluted gas. This mass increase is due to the gravitational field of the central black hole that compresses both fermion systems, and the accumulation of energy in the Fermi spheres of electrons and protons is at the expense of gravitational energy. 
 
The energy of the gravitational field  accumulates itself in Fermi spheres of electrons and protons in a continuous way during matter compression and then is suddenly released at passing the rim of the photon sphere.  The ratio of total Fermi sphere energies of electrons and protons is $\frac{E_n(r^*)}{E_p(r^*)}\simeq 1.4$. The Fermi energy of electrons with Fermi momentum $p_F(r^*)=5.48 \times  10^{-19}$ kg m/s equals to $\varepsilon_F= 1 $ GeV (it is the upper possible energy of emitted photons), which in thermal scale (in units at $k_B=1$) is of order of $9 \times 10^{12}$ K -- this makes the electron liquid quantumly degenerated at lower temperatures (quasars are not source of thermal gamma radiation \cite{shapiro}, thus their actual temperatures are much lower). The Fermi energy of protons with Fermi momentum $p_F(r^*)=5.48 \times  10^{-19}$ kg m/s equals to $\varepsilon_F= 0.4 $ GeV (it is the upper possible energy of emitted photons by jumping of protons), which in thermal scale (in units at $k_B=1$) is of order of $4 \times 10^{12}$ K.
 
The release  of energy  due to the  collapse of the  Fermi sphere of charged particles
undergoes according to the Fermi golden rule scheme for quantum transitions \cite{landau1972}, when such transitions are admitted by the local revoking of Pauli exclusion principle (cf. Appendix \ref{b}).  Charged carriers (electrons and protons) couple to the electromagnetic field and the matrix element of this coupling between an individual particle state in the Fermi sphere and its ground state is the kernel of the Fermi golden rule formula for transition  probability per time unit for this particle.
This interaction depends also on electromagnetic field strength (it arises from the single-particle kinetic energy with momentum $\mathbf{p}$ substituted by the kinematic momentum $\mathbf{p}\pm e\mathbf{A}(\mathbf{r},t)$ and developed to linear term with respect to $\mathbf{A}$, which is the vector potential of the electromagnetic wave at gauge that $div\mathbf{A}=0$, $\pm$ corresponds to proton and electrons, respectively), thus the increasing number of excited photons strengthens the coupling (via the increase of  $\mathbf{A}$) in the similar manner as at stimulated emission (known from e.g.,  laser action) and accelerates quantum transition of the Fermi sphere collapse.

Note that the above energy estimation has been done in conventional rigid coordinates, time and spherical coordinates  $t,r,\theta,\phi$ of the remote observer. The Schwarzschild metric written in these coordinates has the form \cite{schwarzschild},
\begin{equation}
	\label{metryka1}
	-c^2d\tau^2=-\left(1-\frac{r_s}{r}\right)         	c^2dt^2+\left(1-\frac{r_s}{r}\right)^{-1}dr^2+r^2(d\theta^2+sin^2\theta d\phi^2),
\end{equation}
where $\tau$ is the proper time, $t$ is the time measured infinitely far of the massive body, $r_s=\frac{2GM}{c^2}$ is the Schwarzschild radius ($G$ is the gravitational constant, $c$ is the light velocity in the vacuum). The Schwarzschild metric has a singularity at $r=0$, which is an intrinsic curvature singularity. It also  has a singularity on the event horizon $r=r_{s}$ due to the second term in (\ref{metryka1}). The metric (\ref{metryka1}) is therefore defined  on the exterior region $r>r_{s}$ or on the interior region $r<r_{s}$.  However, the singularity on the event horizon disappears, as one sees in other  coordinates.  At passing the event horizon  time-type and space-type intervals mutually change their role, which is also an artifact related to the choice of ordinary remote observer coordinates in Scwarzschild metric. For $r\gg r_{s}$, the Schwarzschild metric is asymptotic to the standard Lorentz metric on Minkowski space. The Schwarzschild metric is a solution of Einstein field equations in empty space for a non-rotating and uncharged spherical body, meaning that it is valid only outside the gravitating body with the mass $M$. That is, for a spherical body of radius $R$ the solution is valid for $ r>R$. To describe the gravitational field both inside and outside the gravitating body the Schwarzschild solution must be matched with some suitable interior solution at $r=R$ such as the interior Schwarzschild metric \cite{frolov}. In the case of a classical concept of a black hole, $R=0$ and the above described problem disappears.
 
The singularity at $r = r_s$ divides the Schwarzschild coordinates in two disconnected patches. The exterior Schwarzschild solution with $r > r_s$ is the one that is related to the gravitational fields of stars and planets. The interior Schwarzschild solution with $0 \leq r < r_s$, which contains the singularity at r = 0, is completely separated from the outer patch by the singularity at $r = r_s$. The Schwarzschild coordinates therefore give no physical connection between the two patches, which may be viewed as separate solutions. The singularity at $r = r_s$ is an illusion however; it is an instance of what is called a coordinate singularity. As the name implies, the singularity arises from a choice of coordinates or coordinate conditions. When one changes  to a different coordinate system (for example Lemaitre, Eddington–Finkelstein, Kruskal–Szekeres, Novikov or Gullstrand–Painlevé coordinates \cite{novikov,kruskal,szekeres}) the metric becomes regular at $r = r_s$ and can extend the external patch to values of $r$ smaller than $r_s$. Using a different coordinate transformation one can then relate the extended external patch to the inner patch. The queer property of Schwarzschild metric is related with the fact that it is impossible to describe the outside and inside of a black hole's event horizon simultaneously in the same rigid and stationary (time independent) metric \cite{lanfield}. The Schwarzschild metric belongs to such a class and it describes the outside of the event horizon using an ordinary coordinate system of the remote observer. In this metric the inside of the horizon is inaccessible (its volume is zero in this metric) and each matter movement terminates for $t\rightarrow \infty$ at the event horizon. If to change, however, to other coordinates like Kruskal-Szekeres or Novikov and to exchange the time $t$ by the proper time $\tau$, then matter smoothly passes the event horizon within a finite proper time period and terminates any movement (also within a proper time period) in central singularity. Different properties of various metrics for the same folded spacetime illustrate different slicing of the same four-dimensional curvature into its space and time components possible to be done in various ways,  with emphasizing of Kruskal-Szekeres metric, being the maximally extended solution of the Einstein equations (analytic in the whole accessible domain) \cite{kruskal,szekeres}.

Nevertheless, for the purpose of the present paper we consider only the upper vicinity of the event horizon in which the Schwarzschild metric well describes geodesics in terms of conventional rigid and stationary  coordinates. They are certainly convenient at the critical innermost unstable circular radius $r=1.5 r_s$. Hence, even if in  Eq. (\ref{fmm}) one substitutes $dV(r)$ by the proper volume at the distance $r$ from the central singularity, which  according to the Schwarzschild metric (\ref{metryka1}) is
\begin{equation}
	\label{correction}
	d{\cal{V}}=\left(1-\frac{r_s}{r}\right)^{-1/2}drr^2 sin\theta d\theta d\phi,
\end{equation}
the change  is only by the factor $\left(1-\frac{r}{r_s}\right)^{-1/2}$. $d{\cal{V}}$ is by this factor  greater than $dV=drr^2sin\theta d\theta d\phi$ observed by the remote observer. At $r^*=1.5r_s$ this factor is ca. 1.7, which gives the reduction  of $p_F$ caused by the gravitational curvature by factor ca. $1.7^{-1/3}\simeq0.84$, which does not change orders in the  estimations presented in this paragraph. The change of $p_F$ by one order of the magnitude would need the closer approaching the Schwarzschild horizon, at $r\simeq 1.000001 r_s$, i.e., rather distant from the $r^*=1.5 r_s$. Hence, for the rough estimation of the effect of Fermi sphere collapse  the correction (\ref{correction})
is unimportant and can be included as the factor $0.84$ to the right-hand side of Eq. (\ref{fmm}), which does not change the orders in the energy estimation.
 
\label{efficiency}
 
\section{Supplement to the conventional model of accretion disc and comparison with observations}
 
The collapse of Fermi spheres in  degenerate Fermi systems  at passing  the innermost unstable circular orbit in Schwarzschild metric  (photon sphere rim of a nonrotating  black hole) does not conflict with the conventional models of the accretion disc of quasars \cite{sunaev,novikov}. The latters base on Shakura-Sunyaev classical hydrodynamic approach to plasma in accretion disc \cite{sunaev} where the inverse transfer of the orbital momentum is modelled by friction and turbulence factors allowing  to increase plasma internal energy on the cost of the black hole gravitational energy \cite{salpet,zeld,ksiazka}.  Then the hot plasma irradiates energy by the thermal radiation \cite{sunaev,merloni}. This gives, however, only soft photons, and to elucidate giant luminosity of quasars (or micro-quasars) in X-ray range, the mechanism of inverse Compton scattering of soft photons on hot electrons or ions is invoked \cite{shapiro}. The model of hot accretion disc, originally proposed for micro-quasar Cignus X-1 with 15 Sun mass black hole \cite{shapiro} assumes extremely high temperature of the inner part of the accretion disc ($10^9$ K for electrons and $10^{11}$ K for ions) to assure sufficient energy of charge carriers needed to Comptonization of soft photons (the identification of a sufficiently abundant soft photon source is not clear, however). Though the Comptomization mechanism in this hot accretion disc model is adjusted to X-ray radiation luminosity of Cignus X-1, the generalization of the model to extremely luminous giant quasars with supermassive black holes ($\sim 10^9$ Sun mass or larger) \cite{supereddington} is problematic, since the temperature $10^9 - 10^{11}$ K  of hot plasma in vast disc seems to be unrealistic. Nevertheless, some numerical simulations of developments of  Shakura-Sunyaev and Thorne-Novikov  model \cite{sunaev,novikov}  done recently by Fragile et al. \cite{6p} allow to match to observable luminosities of some not distant ($z<0.3$) active binary black hole objects \cite{farahzest}, but rather not to super-luminous remote quasars \cite{supereddington}. Higher radiation efficiency has been modelled within conventional classical magneto-hydrodynamic approach by inclusion of hypothetical giant magnetic component to accretion plasma in the case of a spinning black hole \cite{24p}, assuming, however, unrealistic extremal accretion mass rate to gain sufficiently large luminosity.
 
All hydrodynamic or magnetohydrodynamic models of matter accretion onto a black hole are applicable, however, only relatively far from the event horizon of the black hole. In all such models the inner edge of the accretion disc is assumed to be located well above the photon sphere of the black hole (the latter coincides with radius of the innermost unstable circular orbit in Schwawrzschild metric, $1.5 r_s$, $r_s= \frac{2GM}{c^2}$), i.e., even more distant than the innermost stable circular orbit with the radius $3r_s$ (conventionally assumed the inner edge  of the accretion disc is at $\sim 6 r_s$ \cite{shapiro}).
 
The collapse of the Fermi spheres described in the present paper takes place at the rim of the photon sphere at $r=1.5 r_s$, in the region completely neglected in conventional hydrodynamic models of matter accretion \cite{sunaev,novikov,6p,24p}, thus this mechanism does not interfere with thermal (including bremsstrahlung) and Comptonization mechanisms for radiation emission from more distant parts of the accretion disc (typically for $r>6 r_s$). Inclusion of the quantum effect of Fermi sphere collapse at the photon sphere rim ($r=1.5 r_s$) can, however, considerably  supplement developments of Shakura-Sunyaev-Thorne-Novikov  classical hydrodynamics approach \cite{sunaev,novikov} not applicable in close vicinity of the event horizon. The release of high energy photons (depending on the accretion mass rate and the black hole mass and governed by the Fermi momentum in compressed plasma) due to Fermi sphere collapse can add to the total luminosity of quasars and micro-quasars, allowing for the avoidance of some parameter-fitting  problems and shortages of conventional classical models \cite{6p,24p}.
 
The rejection of the quantum statistics in a system of indistinguishable identical particles at passing the rim of the photon sphere is a general property of any black hole, regardless of its mass.
The energy per unit volume stored  in the Fermi sphere of a degenerate Fermi system is a monotonic function of the density of matter. Its maximum is attained for the uppermost possible density of compressed electron-hadron plasma, as in extremely bright quasars or in the case of neutron star mergers exceeding the Tolman-Oppenheimer-Volkoff  stability limit (the uppermost density of compressed Fermi liquid is of order of the density of hadrons in atom nuclei).
In the case of quasars the electron and proton Fermi sphere collapse in the stream of ionized matter in accretion disc  gives the steady luminosity $\sim 10^{40}$ W for  $\sim 10^9$ Sun mass supermassive black holes consuming ca. 10 Sun mass per year ($0.1$ Earth mass per second) for a long time. In the  case of a neutron star merger which exceeds Tolman-Oppenheimer-Volkoff limit of ca. $2.3$ Sun mass compressed to the uppermost density of hadrons, the merger rapidly collapses  due to the relief of internal pressure caused by local recall of the Pauli exclusion principle. The released energy due to the collapse of neutron Fermi sphere in the neutron star merger reaches  $10^{47}$ J (this energy  partly escapes from the photon sphere of a rising black hole in the form of a short giant burst of gamma-rays). In both these extremal cases of the matter compression (in super-luminous quasars and at collapse of neutron star mergers)  the efficiency of mass to radiation conversion is ca. 30 \%, not achievable in any other physical mechanism except for matter-antimatter annihilation (the efficiency of nuclear fusion in stars is only of order $0.7$ \% for mass to energy conversion rate).
 
In the case of not extremal matter compression in the accretion disc of a black holes (as in many active galactic nuclei or in micro-quasars with lower rate of the matter influx) the radiation emitted due to Fermi sphere collapse is less intensive (the mass to energy conversion rate is lower than 30 \%) and softer,  but still contributes to the total luminosity and can help to explain radiation properties of observed binary black hole systems \cite{farahzest,supereddington}. In particular, the Fermi sphere collapse can help to elucidate the relatively short-lasting  brightening of active galactic nuclei, like the recently observed for AGN 1ES 1927+654 \cite{ga1flare}. The 100-fold increase of its luminosity within a few months period would be associated with accidental increase of the matter consumption rate during the corresponding time. If this accidental matter influx is not extremal, then photons emitted due to the Fermi sphere collapse  may not reach over-MeV energy (cf. Table \ref{tab-kw1}) and are not able to produce electron-positron pairs in the ergosphere of this spinning black hole. However, the massive isotropic flux of lower energy photons caused by the Fermi sphere collapse  of electrons and protons may push electron-positron pairs created in the ergosphere according to Blandford-Znajek electromagnetic mechanism \cite{znajek} towards the event horizon, lowering in this way their evaporation to jets across ergosphere nodes. The model by Blandford-Znajek \cite{znajek} gives the theory of jet formation for  spinning Kerr-like black holes where due to dragging  of reference frame in Kerr metric the magnetic field carried with the accretion matter  rotates. The rotation of the ergosphere causes the magnetosphere inside it to rotate, the outgoing flux of angular momentum results in extraction of energy from the black hole. The magnetic field beams in the form of jets and electrons and positrons diffuse across nodes of the ergosphere and next are  highly accelerated in a magnetic field beam in jets producing intensive X-ray radiation. The hypothetical source of electron-positron pairs is a strong electrical field created by the rotating magnetic field frozen in the ergosphere.  The sufficient intensity of an electric field to generate particle-antiparticle pairs in the ergosphere is, however, speculative. A collapse of Fermi sphere in plasma approaching the Kerr black hole would be helpful here, as supplying the abundance of over-MeV photons in the case of extreme matter influx, when the energy of these released photos can reach GeV level and can produce large number of electron-positron pairs, which are able to power up jets besides the Blandford-Znajek mechanism. Nevertheless, in the case when the released photons are sub-MeV at smaller rate of matter consumption by a black hole, then they cannot produce  additional electron-positron pairs inside the ergosphere but can push towards the event horizon those created according to Blandford-Znajek mechanism, reducing their supply to jets.    	This could explain   the temporal change in the radiation spectrum during brightening episode consisting  in optical 100 fold increase of the luminosity and simultaneous lowering of X-ray radiation  (the latter probably due to reduction of the amount of electrons and positrons in the jet, the source of X-ray radiation in jets), without need to speculate on remagnetisation  of the AGN and quenching of its jets by oppositely magnetized gas cloud  during this episode \cite{ga1flare}. When the Fermi energy in accreting plasma does not exceed MeV, then the collapse of electron and proton Fermi spheres can produce an increase of the luminosity with lower frequency (just as has been observed for AGN 1ES 1927+654) and, simultaneously, can temporarily reduce intensity of X-ray radiation form jets, also in agreement with observations.

\section{Conclusions}
 
We proposed a supplementary model for conversion of gravitational energy  close to a giant black hole of quasar into electromagnetic radiation with the efficiency reaching ca. 30\% of mass to energy conversion rate as observed in super-luminous quasars.  This so effective energy engine is quantum in the nature and is based on	the collapse of the Fermi spheres of electrons and protons in dense plasma flow approaching in a spiral way the Schwarzschild horizon near the photon sphere of a giant black hole of the quasar. The collapse of electron and proton Fermi spheres takes place due to the restriction imposed on particle classical trajectories close to the Schwarzschild zone, beneath  the innermost unstable circular orbit (the rim of the photon sphere of non-rotating and uncharged black hole). In this region the homotopy of trajectories qualitatively changes. Below the photon sphere rim (coinciding with sphere with radius of the innermost unstable circular orbit, $r=1.5r_s$ where $r_s$ is the Schwarzschild radius) only existing trajectories unavoidably one-way spiral onto Schwarzschild horizon for particles passing inward the photon sphere. This precludes particle exchanges in this region and thus locally revokes Pauli exclusion principle, which results in the collapse of Fermi spheres of electrons and protons.  The related release of the energy stored in Fermi spheres of  degenerate quantum liquids of electrons and protons has a continuous character provided a steady supply of galactic gas to the accretion disc, and can contribute substantially to the luminosity of quasars for a long term.

The described mechanism of Fermi sphere collapse below the innermost unstable circular orbit of a black hole seems to be universal, but the energy efficiency depends on  the supply of the matter to be consumed by the black hole and is limited by the uppermost density of matter compressed by the gravitational singularity at the rim of the photon sphere. The uppermost density agrees with the limiting density of fermions like in atom nuclei or in neutron stars achieving the Tolman-Oppenheimer-Volkoff stability limit. The collapse  of a neutron star merger, which has exceeded the limit of stability, may be also accompanied by giant emission of photons released during the rapid decay of the Fermi sphere of neutrons, when neutrons not compressed by quantum degeneracy pressure decay into protons and electrons interacting with the electromagnetic field. The energy stored in the neutron Fermi sphere of neutron star achieving Tolman-Oppenheimer-Volkoff limit is of order of $ 10^{47}$ J,  just  the same as the typical energy of frequently observed  short giant gamma-ray bursts, which would be sourced by neutron Fermi sphere decays. Another support for the presented model may be associated with the mechanism of creation of jets of rotating quasars. The model by Blandford and  Znajek of jet formation would be supplemented with new mechanism of the creation of electron-positron pairs in the ergosphere of a Kerr black hole by GeV$-$MeV photons released here in large amount in due of collapse of Fermi spheres of electrons and protons in analogy to the process described for nonrotating Schwarzschild black hole. In the case of smaller portion of mass consumption by a black hole, the decay of Fermi spheres in falling matter does not achieve the highest efficiency but can be still helpful in elucidation of temporal  changes in luminosity and spectral changes of less intensive radiation sources like recently observed incident in AGN 1ES 1927+654. 
 
\appendix
\section{Quantum statistics of indistinguishable identical particles}
 
\label{a}

 A system  of $N$ identical and indistinguishable particles follows its classical dynamics  in multiparticle configuration space, $F_N=(M^N-\Delta)/S_N$, where $M$ is the manifold on which all identical particles are placed (e.g., $R^3$ position space for 4D spacetime or $R^2$ plane for 3D spacetime), $M^N=M\times\dots\times M$ is $N$-fold product of $M$ to account for coordinates of all particles for which the manifold $M$ is equally accessible, $\Delta$ is the diagonal subset of $M^N$ (with points of $M^N$ with coinciding coordinates of at least two particles), subtracted from $M^N$ in order to assure particle number conservation. The division by $S_N$ (permutation group of $N$ elements) introduces indistinguishability of  identical particles, i.e., points in $F_N$ which differ by numbering of particles only, are unified.
 The topological properties of the multiparticle configuration space $F_N$ are represented by homotopy groups of this space  \cite{spanier1966,mermin1979aaa}. The first homotopy group $\pi_1(F_N)$, called as the braid group, collects nonequivalent multi-strand trajectory closed loops in the space $F_N$, joining particle distributions which differ only in particle numbering (such points are unified in $F_N$). These loops describe exchanges of identical indistinguishable particles on which bases the notion of quantum statistics of particles. Scalar unitary representations of the braid group (1DURs, one dimensional unitary representations) define quantum statistics  \cite{mermin1979aaa}.

 Quantum statistics is a result of quantum  indistinguishability of identical particles  and in the case of 4D spacetime (3D position space) scalar unitary representations of the related braid group  lead to   fermionic or bosonic type for particles.  They  differ by antisymmetry or symmetry of  multiparticle wave functions at  exchanges of wave function argument numeration. In the framework of second quantization, bosonic and fermionic statistics are related with the
 commutation or anticommutation of  local field operators of particle creation and annihilation, respectively. It is all a consequence of the rigorous  definition of quantum statistics in collective systems of identical and indistinguishable particles  in terms of  scalar unitary representations of the braid  groups, as quantum statistics is a  non-local collective effect with  roots in  topology  \cite{leinaas1977}.
There exist as many different types of quantum statistics corresponding to classical particles in a system as there are different 1DURs of the braid group for this system.
 
\begin{figure}
	\centering
	\includegraphics[width=0.8\columnwidth]{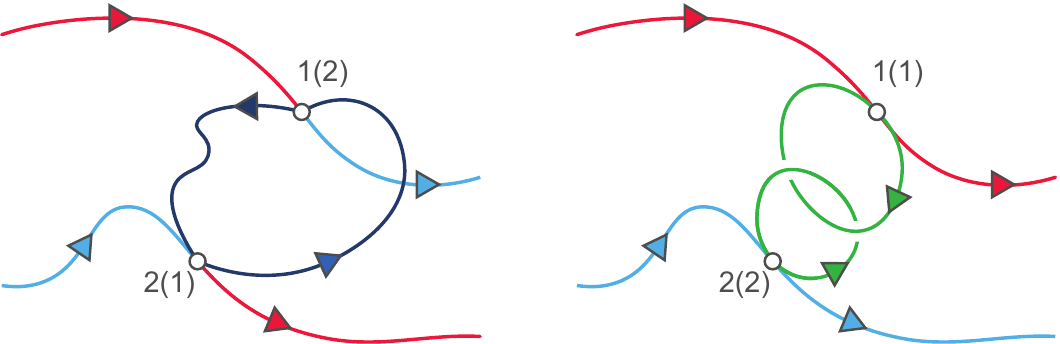}
	\caption{\label{gow11111} To a multiparticle trajectory in the multiparticle configuration space $F_N$ (in the illustration for $N=2$, the configuration space of indistinguishable identical particles is $F_2=(M^2-\Delta)/S_2$) one can add an arbitrary loop from the  braid group $\pi_1(F_2)$ -- two examples of distinct braids attached are shown for the illustration. Due to nonhomotopy of various braids from the braid group (linking positions of particles at some intermediate time instant that may differ by a permutation of particle numeration) the trajectories with various braids attached are topologically nonequivalent, i.e., cannot be transformed one into another by continuous deformations -- they are nonhomotopic.	}
\end{figure}

 Different 1DURs of a particular braid group assign distinct quantum statistics corresponding to the same original classical particles. The linkage of 1DURs of braid groups with quantum statistics can be  clarified  within Feynman path integral formalism. Feynman path integral for a single particle is defined as follows \cite{feynman1964},
\begin{equation}
	\begin{array}{ll}
		I(\mathbf{r}, t; \mathbf{r}', t')=
		\int d\lambda e^{iS[\lambda(\mathbf{r}, t; \mathbf{r}', t')]/\hbar},
	\end{array}
	\label{path0}
\end{equation}
where $           	I(\mathbf{r}, t; \mathbf{r}', t')$ is the so-called quantum propagator, i.e., the matrix element of the quantum evolution operator in the position representation,
$S[\lambda(\mathbf{r}, t; \mathbf{r}', t')]$ is the classical action of the particle along the trajectory $\lambda$ starting from point $\mathbf{r}$ at time instant $t$ and finishing in $\mathbf{r}'$ at time instant $t'$. Integration over the  trajectory space with the measure $d\lambda$ accounts for
contributions of all possible trajectories joining fixed start and final points in the configuration space. The propagator $I(\mathbf{r}, t; \mathbf{r}', t')$ is the complex amplitude
 of the probability for quantum transition between start and final points.
The path integral (\ref{path0}) can be generalized
onto the case of a system of $N$ identical indistinguishable particles.
For such $N$-particle system the path integral has the form \cite{wu,lwitt,chaichian2},
\begin{equation}
	\begin{array}{l}
		I(\mathbf{r}_1,\dots, \mathbf{r}_N, t; \mathbf{r}'_1,\dots, \mathbf{r}'_N, t')\\
		=\sum_{l\in \pi_1(F_N)}e^{i\alpha_l}\int d\lambda_l e^{iS[\lambda_l(\mathbf{r}_1,\dots, \mathbf{r}_N, t; \mathbf{r}'_1,\dots, \mathbf{r}'_N, t')]/\hbar},\\
	\end{array}
	\label{path}
\end{equation}
where $I(\mathbf{r}_1,\dots, \mathbf{r}_N, t; \mathbf{r}'_1,\dots, \mathbf{r}'_N, t')$ is the propagator for the total $N$-particle system, i.e., is the matrix element of the quantum evolution operator of whole multiparticle system  in position representation, which determines the probability amplitude (complex one, in general) for quantum transition between  a start point $(\mathbf{r}_1,\dots, \mathbf{r}_N)$ in the multiparticle coordination space $F_N$ at time instant $t$  and a final point $(\mathbf{r}'_1,\dots, \mathbf{r}'_N)$ in the space $F_N$ at time instant $t'$. Nevertheless, because of particle indistinguishability the numbering of particles can be changed in an arbitrary manner during the way between start and final points in $F_N$. This  can be imagined as adding an arbitrary braid loop  to arbitrary intermediate point of $N$-strand open trajectory in $F_N$ linking start and final points in path integral (as illustrated schematically in Fig. \ref{gow11111}). The addition of an arbitrary finite number of braids to some path in path integral results in the attachment of only one braid, which is the group product of all added braids. Braid groups are countable or finite, thus the index $l$ in Eq. (\ref{path}) is discrete.
 
 Different loops from the braid group are nonhomotopic, i.e. they cannot be transformed one into another by any continuous deformation without cutting, hence the whole space of paths of $N$ particles (the domain of the path integral)	decomposes into subdomains numbered by elements of the braid group (which can be indexed by $l$ in Eq. (\ref{path})). These subdomains are disjoint and nonhomotopic, thus it is impossible to define a common  uniform measure for path integration over the whole domain (because of  discontinuity between its sectors). Instead the separate measures of trajectories in particular sectors must be defined  and
in Eq. (\ref{path}) they are represented by $d\lambda_l$ measures for integration   numbered by the $l$-th element of the braid group $\pi_1(F_N)$ (braid groups always are  countable or finite because they are generated by the finite number of generators, cf. e.g., \cite{birman}, hence the index $l$ is discrete). $S[\lambda_l(\mathbf{r}_1,\dots, \mathbf{r}_N, t; \mathbf{r}'_1,\dots, \mathbf{r}'_N, t')]$ denotes the classical action for the trajectory $\lambda_l$ joining start and final points in the configuration space $F_N$ with the $l$-th braid loop attached to initially open loopless trajectory. The contributions of various sectors of the trajectory domain must be added up in the final formula for the propagator (\ref{path})  with some arbitrary but unitary weight factor $e^{i\alpha_l}$ (the unitarity follows from the causality constraint of quantum evolution). These unitary factors (the weights for various sectors of the trajectory domain) $e^{i\alpha_l}$ in Eq. ({\ref{path}), form a one-dimensional unitary representation (1DUR) of the  braid group \cite{lwitt}. Distinct unitary weights in the path integral (\ref{path}), i.e., distinct 1DURs of the braid group, determine different types of quantum particles corresponding to the same classical ones. Because braids describe particle exchanges on manifold $M$,  thus  1DURs of the braid group assign quantum statistics in the multiparticle system.
	
1DURs of braid groups depend on the homotopy class of trajectories in $F_N$. For $N\geq2$ and $dim M\geq3$  the braid group always is equal to the permutation group,  $\pi_1(F_N)=S_N$. The permutation group $S_N$ has only two different 1DURs,
\begin{equation}
	\label{3d}
	\sigma_i\rightarrow \left\{
	\begin{array}{l}
		e^{i0},\\
		e^{i\pi},\\
	\end{array} \right.
\end{equation}
defining either bosons (for $e^{i0}$) or fermions (for $e^{i\pi}$). In Eq. (\ref{3d})  $\sigma_i$ with $i=1,\dots,N-1$ are the generators of the braid group (here the permutation group), i.e., they define the elementary exchanges of  $i$-th particle with $(i+1)$-th one, at some fixed  but arbitrary numbering of all $N$ particles in the system. For three dimensional manifolds $M$ (or of higher dimension) $\sigma_i^2=\varepsilon$ (neutral element), which causes that the braid group  coincides in this case with the  permutation group with $N!$ elements and only two different 1DURs.

For two dimensional manifolds $M$  the braid group is not equal to $S_N$ and instead is infinite highly complicated countable group \cite{mermin1979aaa,birman} (braid groups are  multi-cyclic groups  generated by finite number of generators $\sigma_i$, hence are countable groups). For two dimensional manifolds $M$, $\sigma_i^2\neq \varepsilon$, which causes a difference with respect to the permutation group. For $M=R^2$ the braid group was  described originally by Artin \cite{artin1947}. The Artin group has an infinite number of 1DURs in the form  $\sigma_i\rightarrow e^{i\alpha}$, $\alpha \in[0,2\pi)$. The related distinct quantum statistics correspond to various $\alpha$ and are referred  to so-called anyons  \cite{wilczek} and represent  fractional statistics besides fermionic or bosonic ones (cf. also Appendix \ref{pauli}).

The above demonstrates  that quantum statistics is conditioned by the topology constraints which govern over the  homotopy of trajectories in  $F_N$, classical multidimensional configuration space  of identical indistinguishable particles. Additionally, there are evidences (including experimental verification) that quantum statistics of the same classical particles can be even changed by external topology-type factors. An example of such a behaviour is the fractional quantum Hall effect, when the strong quantizing magnetic field perpendicular to a planar system of interacting electrons can confine the trajectory class by the cyclotronic effect.
This causes  the modification of the structure of the braid group and next of its 1DURs.  Repulsing electrons on a planar manifold are uniformly distributed  at $T=0$ K and form a regular triangular (hexagonal)  Wigner-type planar  classical lattice. As braids are multi-strand trajectories which exchange positions of particles, they can be defined exclusively in the case when braid sizes fit perfectly to positions of electrons in this lattice. However, the perpendicular magnetic field causes in 2D a finite size cyclotron trajectories and the braids are of the similar finite size as no other trajectories exist at magnetic field presence. Therefore, the braids and related statistics can be assigned only if cyclotron orbits and braids are commensurate with the lattice-type distribution  of electrons. Inclusion of the possibility for matching by multiloop braids also next-nearest neighbouring electrons besides the nearest ones in the electron distribution results in the  hierarchy of filling rates of Landau levels (discrete single-particle quantum states at magnetic field presence), which perfectly elucidates experimentally observed fractional quantum Hall effect hierarchy \cite{annals2021,pan2003,laughlin2}.
The above example shows that the quantum statistics is flexible to trajectory topology changes induced  by the magnetic field in  2D multielectron system and also by electrical field vertically applied to multilayer Hall systems (in bilayer graphene the vertical electrical field can block inter-layer tunnelling of electrons, which precludes hopping of trajectories between layers and eventually changes the statistics of carriers, what is visible in the experiment reported in \cite{maher}). Fractional quantum Hall effect demonstrates that quantum statistics of the same classical electrons can vary in response to external topological type factors, like external magnetic or electrical fields.

In 3D spatial manifolds, the fractional Hall effect does not exist since cyclotron braids in 3D have an arbitrary size (because of a possible drift motion along the magnetic field direction). However, some other topological constraints on braid groups and their unitary representations can be imposed in the case of 3D manifolds. In the present paper we prove that the extremely strong gravitational field close to the Schwarzschild event horizon can restrict trajectory homotopy, which completely prohibits the braid group organization and locally washes out quantum statistics.

Finally let us also comment on quantum statistics in terms of multiparticle wave functions satisfying the Schr\"odinger equation.
The braid groups and their unitary scalar representations define also quantum statistics expressed in terms of the multiparticle wave function $\Psi(\mathbf{r}_1,\dots ,\mathbf{r}_N)$ of $N$ identical indistinguishable particles in full equivalence to path integral quantization approach.  The arguments $\mathbf{r}_1, \dots, \mathbf{r}_N$ of the wave function $\Psi$ are in fact classical coordinates of all particles on the manifold $M$. If these positions are exchanged along a particular braid from the braid group $\pi_1(F_N)=\pi_1((M^N-\Delta)/S_N)$, then  the wave function $\Psi$ must acquire the phase factor equal to the 1DUR of this braid, as was proved by Imbo and Sudarshan \cite{sud,imbo}. Let us emphasize that the exchanges of coordinates are  not permutations in general, and the paths of exchanges are important, unless the manifold $M$ is a three or higher-dimensional space without linear topological defects, as e.g.,  strings \cite{birman,sud}. In each case the assignment of quantum  statistics for particles on arbitrary  dimension manifold $M$  needs, however,  the possibility to implement a braid group, which is conditioned by the existence of classical trajectories for exchanging particle positions on the manifold $M$. If  trajectories for particle exchanges are prohibited, then the quantum statistics cannot be defined. Such a situation we encounter close to the event horizon  of a black hole, beneath the sphere with radius of the  innermost unstable circular orbit in Schwarzschild metric (Appendix \ref{b}).

\section{Pauli theorem on the connection between statistics and spin}
\label{pauli}
To quantum statistics is addressed also the  famous   theorem by Pauli on spin-statistics connection \cite{pauli}. This theorem  asserts  that quantum statistics of particles with half spin must be of fermionic type, while of particles with integer spin -- of bosonic type. This theorem is  supported by the quantum relativistic reasoning that within  the Dirac  electrodynamics for by spinor described particles the Hamiltonian formulation is admitted for simultaneously particles and antiparticles and to assure positively defined kinetical energy of free particles,  the field operators defining particles (or antiparticles) must  anticommute, thus are of fermionic type \cite{landaushort}. Such a proof is confined, however, to free particles and to three dimensional position space -- the manifold on which particles are located. A wide discussion of  Pauli theorem on spin-statistics connection including various trials of its proof are presented in \cite{duck1,duck2}, where also a rigorous  proof of this theorem in the case of  noninteracting particles is formulated by Duck and Sudarshan. Pauli theorem holds, however, independently of interaction and thus is maintained for arbitrary strongly interacting particles. This is visible  in terms of topology as the actual proof of Pauli theorem must invoke to homotopy type reasoning in view of braid group based quantum statistics definition.   Pauli theorem  follows rather  from the coincidence of unitary irreducible representations of the rotation group which define quantization of spin or angular momentum  \cite{rumerfet} with unitary representations of braid groups nominating quantum statistics \cite{mermin1979aaa}. The agreement between unitary representations of both groups arises  due to the overlap of some elements of the braid group and the rotation group (\cite{jac}, paragraph 3.2.5). The representations which are uniform on group generators must thus agree for both groups. The half spin representation of the rotation group  must agree  with the odd representation  $e^{i\pi}=-1$ of the braid group for 3D manifold (and fermions),  whereas the integer angular momentum representation of the rotation group must agree with even  $e^{i0}=1$ representation of the permutation group (and bosons). Such an approach allows simultaneously for  the extension of Pauli theorem onto 2D manifolds with  anyons (which, in general are neither fermions nor bosons). For 2D position space the rotation group is Abelian, which causes that spin in 2D is not quantized but perfectly agrees with the  continuous  scalar unitary representations of the Artin braid group defining anyon fractional statistics. Moreover, the sketched above topological proof of  Pauli theorem is immune to the particle interaction.

To be more specific, let us note that for 3D manifolds, the rotation group  $O(3)$  has  the  covering group  $SU(2)$  and  the  irreducible unitary representations of $SU(2)$ fall into two classes assigning integer and half-integer angular momenta. These two classes agree with only two  possible scalar unitary representations of the permutation group $S_N$, which is the  braid group for 3D manifolds. The representations of both groups coincide as they have some common elements.  However, for 2D manifolds the rotation group $O(2)$  is Abelian and isomorphic with $U(1)$ group possessing just the same continuous  unitary representations $e^{i\alpha}$, $\alpha\in[0,2\pi)$,  as the Artin group, which is the braid group for $M=R^2$. Thus, in two dimensional space   Pauli theorem also holds  for not quantized spin assigned by $s=\frac{\alpha}{2\pi}$ and similarly continuously changing anyon statistics defined by $e^{i\alpha}$ numbered by  $\alpha\in[0,2\pi)$.

Quantum  statistics and spin, though coincide via the agreement between unitary representations of rotation and braid groups, are in fact independent to some extent, and one can imagine a situation when the spin is still defined but the statistics not, as  in the case of the absence of a braid group. Such a situation occurs in an extremely strong gravitational field inside the black hole beneath the event horizon and also beyond the event horizon but beneath the photon sphere,   as will be demonstrated  in  Appendix \ref{b}.

\section{Homotopy of particle trajectories  in the vicinity of the event horizon in Schwarzschild metric}

\label{b}

Schwarzschild metric \cite{schwarzschild}  describes non-rotating and uncharged  classical black hole with the mass $M$. The folded geometry of the spacetime induced by a point-like mass $M$ can be expressed as the line element for the proper time shift,  $ds=cd\tau$,  given by Eq. (\ref{metryka1}). Schwarzschild radius $ r_s = \frac {2GM} {c^2} $ defines the black hole event horizon, i.e. the boundary surface of an area close to the central singularity  from which neither matter nor light can escape.
	
Additionally, the outer vicinity of the Schwarzschild event horizon is the place where the particle trajectories change quantitatively their geometry resulting in the specific homotopy of these trajectories, which precludes the mutual interchanges of particle positions in some manifold $M^* \subset \tilde{R^3}$ ($\tilde{R^3}$ is the  3D position space $R^3$ folded according to the metric (\ref{metryka1})) for $r\in(r_s,1.5r_s)$. For such a  manifold the multiparticle configuration space $F_N=(M^{*N}-\Delta)/S_N$ becomes simply-connected (as we will demonstrate below).  In such a case the braid group $\pi_1\left((M^{*N}-\Delta)/S_N\right)=\{\varepsilon\}$, i.e., is a trivial group with only neutral element $\varepsilon$ -- the zeroth loop not mixing particle numeration. None exchanges of particle positions are available here, they cannot be defined for particles on the manifold $M^*$. Hence, the statistics of these particles cannot be assigned in this region. For the group $\{\varepsilon\}$, the representation $\varepsilon\rightarrow 1$ (only one possible because $\varepsilon\cdot\varepsilon =\varepsilon$) does not assign bosons, as $\varepsilon$ is not particle exchange. Fermions also cannot be assigned and particles with half spin lose  their statistics including the local revoking of the Pauli exclusion principle for them.

This is a rapid change of the braid group $\pi_1(F_N)$, which outside  the manifold $M^*$, i.e., for outer space beyond the rim of $M^*$, i.e., for $r>r_s$  is  $S_N$ allowing for bosons and fermions there.

Below we will argue that this rapid change of the particle trajectory homotopy takes place at passing the sphere with radius of the innermost unstable circular orbit in metric (\ref{metryka1}), which is at $r=1.5 r_s$ distance from the singularity in $r=0$.

Let us consider trajectories of particles (with mass $m$ vanishingly small in comparison to $M$) in the upper neighbourhood of the Schwarzschild event horizon. These trajectories coincide with geodesics in the metric (\ref{metryka1}).  Because of the spherical symmetry of the gravitational field described by (\ref{metryka1}) these trajectories must lie in planes and without any loss of generality we can consider the geodesic plane $\theta=\frac{\pi}{2}$.
The geodesics for a particle  with the mass $m$ can be determined in various equivalent classical dynamics formulations, e.g.,  by  solution of the Hamilton-Jacobi equation,
\begin{equation}
	\label{geodesics}
	g^{ik}\frac{\partial S}{\partial x^i}\frac{\partial S}{\partial x^i}-m^2c^2=0,
\end{equation}
with $g^{ik}$ metric tensor components corresponding to (\ref{metryka1}) metric \cite{lanfield}.
Eq. (\ref{geodesics}) attains  for the Schwarzschild metric  (\ref{metryka1})  the following form,
\begin{equation}
	\label{jacobi1}
	\left(1-\frac{r_s}{r}\right)^{-1}\left(  \frac{\partial S}{c\partial t}\right)^2-\left(1-\frac{r_s}{r}\right)\left(\frac{\partial S}{\partial r}\right)^2
	-\frac{1}{r^2}\left(\frac{\partial S}{\partial \phi}\right)^2-m^2c^2=0,
\end{equation}
with the function $S$ in the form,
\begin{equation}
	\label{jacobi}
	S=-{\cal{E}}_0t+{\cal{L}}\phi+S_r(r).
\end{equation}
In the above formula  the quantities  ${\cal{E}}_0$ and  ${\cal{L}}$ are the particle energy and its angular momentum, respectively. ${\cal{E}}_0$ and  ${\cal{L}}$ are constants of motion.
Eq. (\ref{geodesics}) can be also applied to define trajectories of photons assuming in (\ref{geodesics}) $m=0$.

 If one  substitutes Eq. (\ref{jacobi}) into Eq. (\ref{jacobi1}) then one can find $\frac{\partial S_r}{\partial r}$. By the integration of this formula one can find,
\begin{equation}
	S_r=\int dr\left[ \frac{{\cal{E}}_0^2}{c^2}\left(1-\frac{r_s}{r}\right)^{-2}-\left(m^2c^2
	+\frac{{\cal{L}}^2}{r^2}\right)\left(1-\frac{r_s}{r}\right)^{-1}\right]^{1/2}.
\end{equation}
The geodesic equation in considered geodesic plane is determined   by the condition $\frac{\partial S}{\partial{\cal{E}}_0}=const.$, which gives radial dependence of the trajectory  $r=r(t)$, and by the condition	$\frac{\partial S}{\partial{\cal{L}}}=const.$, determining the angular dependence $\phi=\phi(t)$                                of the particle trajectory.
The condition $\frac{\partial S}{\partial{\cal{E}}_0}=const.$ gives,
\begin{equation}
	\label{promien}
	ct=\frac{{\cal{E}}_0}{mc^2}\int \frac{dr}{(1-\frac{r_s}{r})\left[\left(\frac{{\cal{E}}_0}{mc^2}\right)^2-\left(1+\frac{{\cal{L}}^2}{m^2c^2r^2}\right)\left(1-\frac{r_s}{r}\right)\right]^{1/2}}.
\end{equation}
The  condition $\frac{\partial S}{\partial{\cal{L}}}=const.$ results in the relation,
\begin{equation}
	\label{faza}
	\phi=\int dr\frac{{\cal{L}}}{r^2}\left[\frac{{\cal{E}}_0^2}{c^2}-\left(m^2c^2+\frac{{\cal{L}}^2}{r^2}\right)
	\left(1-\frac{r_s}{r}\right)\right]^{-1/2}.
\end{equation}

\begin{figure}
	\centering
	\includegraphics[width=0.65\columnwidth]{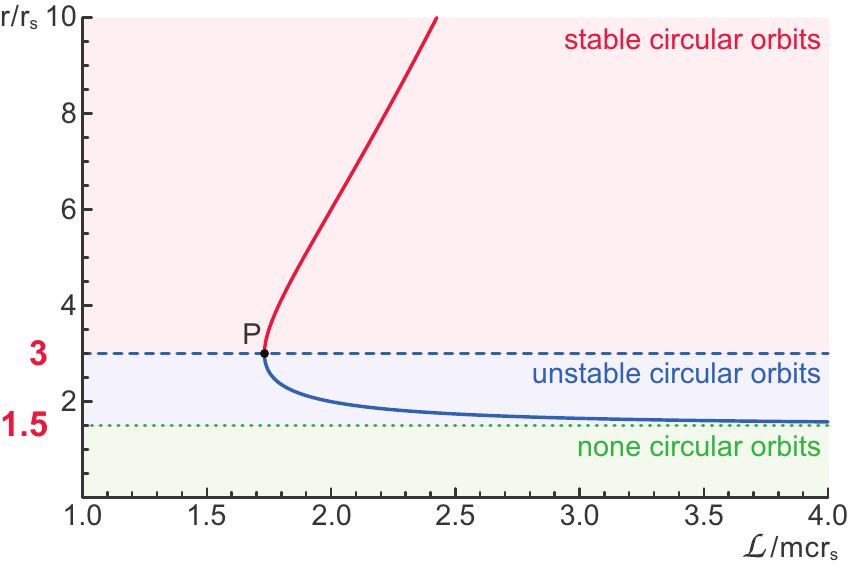}
	\caption{\label{photosphere} Radii of stable (red) and unstable (blue) circular orbits in Schwarzschild geometry. The innermost circular stable orbit occurs at $r=3r_s$ (point $P$ on the level of dashed line in blue colour), whereas the innermost unstable circular orbit occurs at $r=1.5 r_s$ (asymptotic dotted line in green colour). Below $r=1.5r_s$ none circular orbits exist.     }
\end{figure}

Eq. (\ref{promien}) can be rewritten in a differential form,
\begin{equation}
	\label{differential}
	\frac{1}{1-r_s/r}\frac{dr}{cdt}=\frac{1}{{\cal{E}}_0}\left[{\cal{E}}_0^2-U^2(r)\right]^{1/2},
\end{equation}
with the  effective potential,
\begin{equation}
	\label{potencjal}
	U(r)=mc^2\left[\left(1-\frac{r_s}{r}\right)\left(1+\frac{{\cal{L}}^2}{m^2c^2r^2}\right)\right]^{1/2},
\end{equation}
where ${\cal{E}}_0$ and ${\cal{L}}$
are energy and angular momentum of the particle, respectively.

The equation (\ref{differential}) allows for the definition of an accessible region for the motion via the following condition, ${\cal{E}}_0\geq  U(r)     	$.
Moreover, the condition
 ${\cal{E}}_0=  U(r)     	$ defines circular orbits.
Limiting circular orbits can be thus found  by the determination of extrema of $U(r)$.  Maxima of $U(r)$ define unstable orbits, whereas minima stable ones (depending on parameters ${\cal{E}}_0$ and ${\cal{L}}$, which are integrals of the motion). The conditions $U(r)={\cal{E}}_0$ and $\frac{\partial U(r)}{\partial r}=0$ (for extreme) attain the explicit  form,
\begin{equation}
\label{branches}
\begin{array}{l}
	{\cal{E}}_0={\cal{L}}c\sqrt\frac{2}{rr_s}\left(1-\frac{r_s}{r}\right),\\
	\frac{r}{r_s}=\frac{{\cal{L}}^2}{m^2c^2r_s^2}\left[1\pm \sqrt{1-\frac{3m^2c^2r_s^2}{{\cal{L}}^2}}\right],
\end{array}
\end{equation}
where the sign $+$ in the second equation  corresponds to stable orbits (minima of $U(r)$) and the sign $-$ to unstable ones (maxima of $U(r)$). Positions of stable and unstable circular orbits depend on energy ${\cal{E}}_0$ and angular momentum ${\cal{L}}$.
This is illustrated in Fig. \ref{photosphere} -- the upper curve (red one in this figure) gives positions of stable circular orbits (with respect to ${\cal{L}}$) and the lower curve (blue one) gives positions of unstable circular orbits (also with respect to ${\cal{L}}$). The related value of ${\cal{E}}_0$ for each point is given by the first equation of the system (\ref{branches}).

One can notice that the upper curve (the red one)  terminates in  the point $P$ at $r=3r_s$. This point defines   the innermost stable circular orbit. It is  at $r=3r_s$, ${\cal{L}}=\sqrt{3}mcr_s$ and ${\cal{E}}_0=\sqrt{\frac{8}{9}}mc^2$ (point $P$ in Fig. \ref{photosphere}).

The position of the innermost unstable circular orbit is at $r=1.5 r_s$ for ${\cal{L}} \rightarrow \infty $ and ${\cal{E}}_0\rightarrow \infty$ -- it is an asymptotic value defined by the horizontal dotted asymptotic line in Fig. \ref{photosphere} marked  in  green colour. One can note that $r=1.5 r_s$ is also the unstable circular orbit for photons (by taking the limit $m=0$), which defines the so-called photon sphere in Schwarzschild metric. 

Below the radius $r=1.5 r_s$ none circular orbits exist for massive and massless particles  and the corresponding geodesics unavoidably spiral one-way towards the horizon if particles pass this limiting sphere  inwards no matter how high or low the initial energies and angular momenta of  particles are.

At passing the limiting sphere with radius of the innermost unstable circular orbit $r=1.5 r_s$, the qualitative change of trajectories takes place. Beyond this sphere there exist circular orbits but beneath not.
This is in contrast to the classical Newton gravitational singularity, for which circular orbits exist arbitrarily close to the centre.
The difference between the general-relativistic gravitational singularity and the Newtonian one resolves itself next to the absence of conic section
shapes of trajectories below the innermost circular orbit at $1.5 r_s$ in Schwarzschild metric in distinction to Newtonian case where conic section trajectories are possible arbitrarily close to $r=0$. General-relativistic relativistic trajectories below $1.5r_s$ must be only of spiral shape defined by Eqs (\ref{promien})  and (\ref{faza}) for their radius and phase, respectively. This displays the gravitational curvature effect of giant attraction term $\sim - \frac{{\cal{L}}^2}{r^3}$ in effective potential (\ref{potencjal}) below $1.5 r_s$. This term severely limits the shape of  possible trajectories to only short spirals toward the event horizon with a phase shift the smaller the larger $ {\cal{L}} $, which is noticeable from the integral in the equation (\ref{faza}) taken in the range $ (r_s, 1.5r_s) $. Spirals accessible as particle trajectories in the region $r\in (r_s,1.5 r_s)$ can mutually intersect in a single point at most depending on initial conditions of particles at passing  $r=1.5 r_s$, which is in contrast to conic section trajectories accessible for $r>r_s$ allowing  mutual intersection in two points (e.g., section of an ellipse with a circle, hyperbole  or parabola). In the latter case a closed local paths created by two pieces of single-particle orbits in $M$ are possible, in the former case not.

If two particle exchange their positions in $M$, then two inverse and different paths between these particle positions must be available, which together create a closed loop in the manifold $M$. If such loops are impossible as for spirals in the region $r\in(r_s,1.5r_s)$, then particle exchanges are unavailable there.

 The disappearance of circular trajectories in Schwarzschild metric  beneath the sphere of innermost unstable circular orbits   and the obligatory one-way falling down (along a short spirals) onto the event horizon for  particles passing inwards the sphere $r=1.5r_s$, do not allow to construct braids for these particles. None closed  trajectory loops are possible beneath the sphere with radius $1.5r_s$.  This homotopy property is independent of particle mutual interaction as such a local interaction may only locally deform trajectories but cannot change their topological class determined here by the gravitational spacetime curvature. Note additionally, that the homotopy class of trajectories is immune to a change to distinct curvilinear coordinates at the metric of the other choice for the same gravitational singularity. The differences between metrics resolve themselves to various slicing of the invariant folded spacetime into its time and spatial components. In particular, it is impossible to simultaneously describe the outer and inner regions with respect to the event horizon in terms of stationary (time independent) and rigid coordinate system (as of remote observer) \cite{lanfield}. Therefore the Schwarzschild metric (\ref{metryka1}), which is stationary and uses conventional rigid coordinates $(r,\theta, \phi)$, can properly describe the outside of the horizon (with time limit at $ \infty $ for particle motion, i.e., particles spiral onto the event horizon infinitely long
 for the remote observer). Inside the horizon the spatial intervals become time ones and vice versa in metric (\ref{metryka1}), and the spatial volume of this region is zero in Schwarzschild metric. Nevertheless, in nonstationary metrics by e.g., Kruskal and Szekeres \cite{kruskal,szekeres} or Novikov \cite{novikov} (i.e., at time dependent slicing of the spacetime into time and spatial components) the inner region beneath the event horizon is accessible and particles smoothly pass this horizon  in finite proper time period (the proper time is time coordinate in Novikov metric and also is closely linked to time coordinate in Kruskal-Szekeres metric). After passing the event horizon, particles  spiral onto the central singularity and terminate there the movement also within a finite proper time period. The homotopy of trajectories is, however, the same in arbitrary equivalent metrics, precluding particle interchange below the innermost unstable circular orbit. 

Impossibility to exchange particle positions below the innermost unstable circular orbit one can rationalize in the following way.
 Braids describing particle exchanges must be closed loops in the multiparticle configuration space $F_N=(M^N-\Delta)/S_N$ and for the local manifold $M^*$ in the curved space between  the sphere of innermost unstable  circular orbit and any other sphere closer to the horizon,  none such loops exist. This precludes exchange positions of indistinguishable particles in this region because of the absence of suitable single-particle trajectories from which braids must be built. Beyond the innermost unstable circular orbits, single-particle trajectories are of conic section shape (almost the same as in the Newtonian case, though with topologically unimportant additional precession of non-circular orbits, like observable in Mercury orbit around the Sun). Conic section orbits admit local loops for particle pair exchange on manifold $M$ necessary to build generators of the braid group $\pi_1(F_N)$. For Schwarzschild geometry (\ref{metryka1}) such orbits are inaccessible for $M^*$ in the region $r\in(r_s,1.5 r_s)$ and only possible short spirals in $M^*$ do not allow for local particle pair exchanges.

The  stationary Schwarzschild coordinates are convenient to describe in the ordinary rigid coordinates (as in remote system) $t,r,\theta,\phi$ the outer region with respect to the event horizon, thus also the region between the innermost unstable circular orbit and the horizon. In these coordinates any reformulation of path integral (\ref{path}) is not required.  The similar trajectory property as below the innermost unstable circular orbit holds, however, also for the inner of a black hole beneath the horizon, where   all particles unavoidably one-way spiral to the central singularity, and also no other trajectories exist there. This movement can be conveniently parametrized in nonstationary Kruskal-Szekeres or Novikov metrics, though the homotopy class of trajectories is the  same in all curvilinear coordinates. In curvilinear  coordinates the path integral (\ref{path}) must be reformulated e.g., to proper time and Novikov radial coordinate \cite{novikov} or similar non-stationary and non-rigid coordinates in Kruskal-Szekeres metric \cite{kruskal,szekeres}, which, however, does not change the reasoning related to homotopy of trajectories for indistinguishable particles expressed by the braid group. 
For example, in Novikov metric the proper time $\tau$ is the time-coordinate, while the new radial coordinate $R$ is given by the equation,         	
 \begin{equation}
	\label{novikov}
	\frac{\tau}{2M}=(R^2+1) \left[\frac{r}{2M}-\frac{(r/2M)^2}{R^2+1}\right]^{1/2}+(R^2+1)^{3/2}arccos\sqrt{\frac{r/2M}{R^2+1}},
	\end{equation}
(here  $c,G=1$), the angular coordinates $\theta$ and $\phi$  remain unchanged. From (\ref{novikov})
it is evident that the mutual dependence between Novikov radial coordinate $R$ and ordinary radius (measured by remote observer)       	$r$ is time dependent. The change to curvilinear coordinates in new metric does not change the homotopy of trajectories.  		
	
\begin{figure}
\centering
\includegraphics[width=0.7\columnwidth]{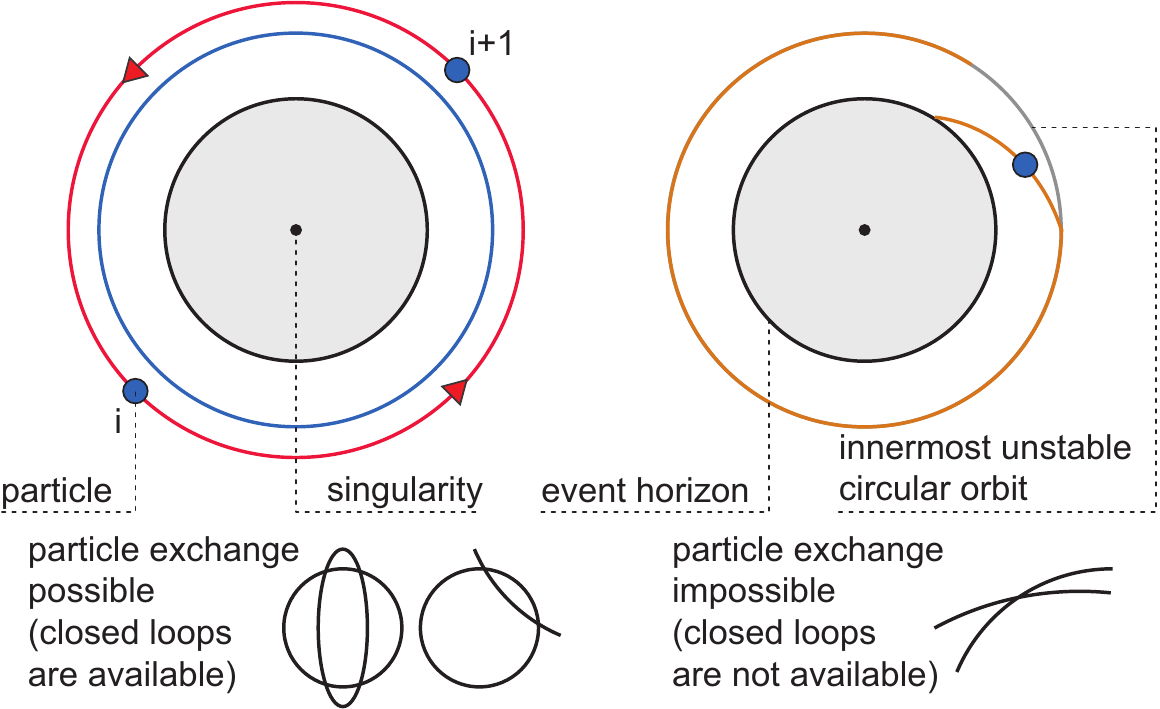}
\caption{\label{gc} Simplified pictorial illustration of the change of trajectory homotopy at passing the innermost unstable circular orbit of a black hole. If circular orbits are available then the particle position interchange is possible, in principle (left picture). When only spiral one-way trajectories are admitted and particles unavoidably fall towards the event horizon (in Schwarzschild coordinates) (central picture) or towards the singularity (right picture), particles cannot mutually interchange  positions. Even if the $i$-th particle  can substitute the position of the other $(i+1)$-th one, the inverse trajectory does not exist no matter how close particles are (central and right pictures).  	}
\end{figure}

To demonstrate the homotopy class of trajectories in the manifold $M^*$ between the innermost unstable circular orbit and the event horizon, the Schwarzschild metric is, however, especially convenient because it does not cause any need to reformulate the path integral (\ref{path}) and the definition of $F_N$. The trajectories in  $F_N=(M^{*N}-\Delta)/S_N$ are parametrized in terms of the ordinary rigid coordinates $(t,r,\theta,\phi)$ and the absence of  montrivial  braid group for this space is apparent below $1.5r_s$.
The elements of a braid group are nonhomotopic classes  of closed loops in $F_N$ assuming indistinguishability of particles.  Because of the division by the permutation group  in $F_N$  definition, this space is not intuitive and differently numbered particle configurations are unified to the same point in $F_N$, which is counter-intuitive. Despite such a limitation of a visualization of braids, two distinct real trajectories in $M$ able to link two particles at some fixed but arbitrary  particle numbering must be available. If such trajectories  are not available, the nontrivial braid group does not exist. The existence of circular orbits assures trajectory topology sufficient to particle interchange -- such trajectories may serve for exchanging of particles located on the same circular orbit and thus, due to indistinguishability, for all particles. Two particles located at ends of an arbitrary  diameter of a circle can exchange their positions along two semicircle trajectories (as illustrated in Fig. \ref{gc}). The existence of circular trajectories (actually, of any closed single-particle trajectory) assures in topological sense the possibility of the braid group implementation.  Note that the closure of circles of single particle orbits does not mean the closure of loops in $F_N$, but some pieces of circular orbits (e.g., semicircles in the example)  allow for the organization of elementary braids (of generators of the braid group, i.e., of elementary exchanges of particles $i$-th with $(i+1)$-th one at some fixed particle numbering, conserving simultaneously positions of the rest of particles \cite{birman,mermin1979aaa}). In other words, the generators of the braid group, $\sigma_i$, $i=1,\dots,N-1$, are $N$-thread trajectory bunches exchanging only $i$-th particle with $(i+1)$-th one when the other particles remain at rest, at some fixed but arbitrary particle numbering. This exchange of $i$-th and $(i+1)$-th particles  must be, however, available in the manifold $M$.    In the case of only on-way directed spiral trajectories   of  particles passing inwards  the innermost unstable circular orbit, the braid loops linking various particles falling onto the horizon cannot be organized. Such trajectories do not admit local closed loops in $M^*$, even if they are deformed  by interparticle interaction. The overwhelming role of the gravitation space-curvature below  the innermost unstable circular orbit is dominating and plays the role of the topological factor in multiparticle systems confining availability of trajectories. This is in contrast to the region beyond the innermost unstable circular orbit, where the presence of circular orbits (and related conic section trajectories) changes  the homotopy class and allows the implementation of the braid group in the form of $S_N$ (as usual for 3D spatial manifolds).  

 Similarly for the inner of the event horizon -- beneath the Schwarzschild surface  the  dynamics of particles is completely controlled by the central singularity and all  particles unavoidably travel here to the singularity along short spirals towards the origin despite any strength of interparticle interaction and initial conditions.  This is visible in Kruskal-Szekeres \cite{kruskal,szekeres} or Novikov \cite{novikov} coordinates. The homotopy of trajectories is immune to the change of curvilinear coordinates in the folded spacetime induced by the gravitational singularity, though trajectories are deformed in their distinct parametrizations.  Also interparticle interaction does not change the  trajectory homotopy though locally can deform trajectories from its free particle shape. Interparticle interaction cannot produce closed cycles from one-way directed spirals predominantly governed by central singularity up to the sphere with the radius of the  innermost  unstable circular orbit. 
The qualitative change of the trajectory homotopy at the innermost unstable circular orbit is schematically illustrated  in Fig. \ref{gc}.

 
\bibliographystyle{JHEP}

\providecommand{\href}[2]{#2}\begingroup\raggedright\endgroup

 

\end{document}